\documentclass[english]{article}
\usepackage[latin9]{inputenc}
\usepackage{amsmath}

\makeatletter

\providecommand{\tabularnewline}{\\}

\usepackage{multirow}
\usepackage{placeins}
\usepackage{babel}

\providecommand{\tabularnewline}{\\}

\usepackage{babel}

\usepackage{babel}

\usepackage{babel}

\makeatother

\usepackage{babel}
\begin{document}
\title{WAVE FUNCTIONS FOR PENTADIAGONAL MATRICES IN THE WEAK COUPLING LIMIT}
\author{Larry Zamick\\
Department of Physics and Astronomy, \\
 Rutgers University, Piscataway, New Jersey 08854}
\maketitle
\begin{abstract}
We consider a pentadiagonal matrix which will be described in the
text. We demonstrate practical methods for obtaining weak coupling
expressions for the lowest eigenvector in terms of the parameters
in the matrix, v and w. It is found that the expressions simplify
if the wave function coefficients are put in the denominator. 
\end{abstract}

\section{Introduction }

Matrix diagonalization is at the heart of the nuclear shell model .
But more than that it is of vital importance in many other branches
, certainly atomic and molecular physics as well as condensed matter.
In nuclear physics we typically use very complicated Hamiltonians
which are often represented as a long string of 2 body matrix elements.
Despite this the output is in some cases tantalizingly simple.In previous
works {[}1-4{]} we studied simple matrices in part for their own sake
but also we tried to connect and gain insight with results of realistic
calculations. We here choose an example from personal experience.
In ref {[}5{]} , using a complicated Hamiltonian we found simple result
that when a binning process was applied the magnetic dipole strength
from the ground state appears to decrease exponentially with excitation
energy. In references {[}1{]} and {[}2{]} we found a similar behavior
when we used very simple tridiagonal matrix indicating there might
be something general about this behavior.Note that although ref {[}5{]}
will have been published after the ``matrix'' papers {[}1-4{]} it
was written earlier and appeared in the archives earlier. It had a clear
influence on these ``matrix'' papers.

.

\section{The Interaction}

In the previous papers the matrices that were addressed included tridiagonal,
pentadiagonal and heptadiagonal{[}1-4{]}.However most of the analytic
work was performed for the tridiagonal case. Here we consider the
more complex pentadiagonal case and address the problem of the ground
state wave functions in the weak coupling limit.

.

We start by showing in Table 1 an (11)x(11) pentadiagonal matrix which
is to also represent a nuclear Hamiltonian. We have dealt with such
a matrix before{[}3{]} but onlt for the case w=v. On the diagonal
we shall consider the case where E$_{n}$=nE.

\noindent\begin{minipage}[t]{1\columnwidth}%
Table 1. An (11) x(11) Pentadiagonal Matrix with 2 Parameters v and
w.%
\end{minipage}

.

\{E0, v, w, 0, 0, 0, 0, 0, 0, 0, 0\},

\{v , E1, v, w, 0, 0, 0, 0, 0, 0, 0\},

\{w, v, E2, v, w, 0, 0, 0, 0, 0, 0\},

\{0, w, v, E3, v, w, 0, 0, 0, 0, 0\},

\{0, 0, w, v, E4, v, w, 0, 0, 0, 0\},

\{0, 0, 0, w, v, E5, v, w, 0, 0, 0\},

\{0, 0, 0, 0, w, v, E6, v, w, 0, 0\},

\{0, 0, 0, 0, 0, w, v, E7, v, w, 0\},

\{0, 0, 0, 0, 0, 0, w, v, E8, v, w\},

\{0, 0, 0, 0, 0, 0, 0, w, v, E9, v\},

\{0, 0, 0, 0, 0, 0, 0, 0, w, v, E10\}\}

.

. We have previously studies such matrices in works by A. Kingan and
L. Zamick{[}1,2{]} and L.Wolfe and L.Zamick {[}3,4{]}, but only for
the case w=v.

\section{Expressions for the lowest eigenvector components of a pentadiagonal
matrix-numerical approach}

..

.Our objective is to obtain the lowest eigenvector of the pentadiagonal
matrix shown above in the weak coupling limit i.e. when both v and
w are much smaller than the energy separations on the diagonal . We
use the word ``obtain'' rather than ``derive''. While a derivation
can be obtained it becomes very complicated. We here offer a practical
method of obtaining the eigenvector components. Of course it is not
diffiicult, with programs like Mathematica to obtain these components
for any strength of the interaction but we are here interested in
the analytical form.We associate the lowest eigenvector wiith a ground
state wave function.

Our method is to choose small values of v and w . We use Mathematica
to obtain the eigenvectors, and fit the eigenvectors components to
a plausible formula. To make things competative we chose the following
set of values:

v = n {*}10$^{-4}$and w = m {*}10$^{-8}$ with m and n small integers.
Why the big difference between v and w? With w we can make a 2 state
jump in one shot but with v we have to make the jump in 2 steps.

By examining the structure of the lowest eigenvalue we find that the
components \{a$_{0}$, a$_{1}$,......,a$_{9}$ , a$_{10}$\} have
the following structure:

\{1, $-v$, (c$_{0}$v$^{2}$+ c$_{1}$w),( d$_{0}$v$^{3}$+ d$_{1}$v
w),( e$_{0}$v$^{4}$+ e$_{1}$v$^{2}$w + e$_{3}$w$^{2}$),......\}.

When the results are present with the numbers in the numerator they
are complicated looking , but in the denominator they are for the
most part integers, and so the formulas look much simpler. Our previous
works shed some light on this, as will be seen in the next section.

.

.

\noindent\begin{minipage}[t]{1\columnwidth}%
Table 2: Components of the ground state wave functions for pentadiagoal
matrices in terms of v and w.%
\end{minipage}

.%
\begin{tabular}{|c|c|}
\hline 
n  & a$_{n}$\tabularnewline
\hline 
\hline 
0  & 1\tabularnewline
\hline 
1  & -v\tabularnewline
\hline 
2  & v$^{2}/2-w/2$\tabularnewline
\hline 
3  & -v$^{3}$/6+vw/2\tabularnewline
\hline 
4  & v$^{4}$/24-v$^{2}$w/4+ w$^{2}$/8\tabularnewline
\hline 
5  & -v$^{5}$/120+v$^{3}$w/12-v w$^{2}$/8\tabularnewline
\hline 
6  & v$^{6}$/720-v$^{4}$w/48+v$^{2}$w$^{2}$/16 -w$^{3}$/48\tabularnewline
\hline 
7  & -v$^{7}$/5040 +v$^{5}$w/240 -v$^{3}$w$^{2}$/48+v w$^{3}$/48\tabularnewline
\hline 
8  & v$^{8}$/40320- v$^{6}$w/ 1440 +v$^{4}$ w$^{2}$/96-v$^{2}$w$^{3}$/96
+w$^{4}$/384\tabularnewline
\hline 
9  & -v$^{9}$/362880+ v$^{7}$w/10080 - v$^{5}$w$^{2}$/945 +v$^{3}$w$^{3}$/288-vw$^{4}$/384 \tabularnewline
\hline 
10  & v$^{10}$ /3628800 -v$^{8}$w/80640+v$^{6}$w$^{2}$/5760 -v$^{4}$w$^{3}$/1152+
v$^{2}$w$^{4}$/768 -w$^{5}$/3840\tabularnewline
\hline 
\end{tabular}

.

.

.

\section{Expressions for the lowest eigenvector components of a pentadiagonal
matrix-analytic approach}

In ref {[}3{]} entitled ``Relation Between Exponential Behavior and
Energy Denominators-Weak Coupling Limit'' L.Wolfe and L.Zamick considered
first a tridiagonal matrix i.e. one for which w=0, and then simpler
pentadiagonl matrix, one for which w=v.

For the tridiagoal case v/E=0.01 wes used. The results were as follows:

.\{0.99995, --0.009999, 0.0000499933,-1.6664{*}10$^{-7}$,4.16592{*}
10$^{-10}$,-8.33171{*} 10$^{-13}$,1.3886{*} 10$^{-15}$, -1.98269{*}
10$^{-18}$,2.47958{*} 10$^{-21,}-2.75506${*} 10$^{-24}$,2.75504{*}
10$^{-27}$\}

These numbers were put in a more suggestive way with a bit of rounding
up.

1,-v/E,(v/E)$^{2}$/2 ,-(v/E)$^{3}$/6 ......... (-1)$^{n}$ (v/E)$^{n}$/n!.

To get a$_{n}$ one has to go the nth order in perturbation theory.
Let H=H$_{0}$+V with H$_{0}$ the diagonal part of the matrix and
V the off diagonal.

$\Psi$= $\varPhi$ + 1/(E$_{0}$-H$_{0}$)QV$\varPsi$ = (1+1/ (E$_{0}$-H$_{0}$)Q
V +1/(E$_{0}$-H$_{0}$) QV 1( E$_{0}$-H$_{0}$)Q V+..... )$\varPhi$

where Q prevents the unperturbed ground state from being an intermediate
state.

To get a$_{n}$to the lowest power in v/E one has to go in n steps-
0 to 1, 1 to 2,...(n-1) to n. In the numerator all the matrix elements
\textless{} (n+1) QVn\textgreater{} are the same, namely v. In
the denominator one gets (E$_{0}$-E$_{1})$.....(E$_{0}$ -E$_{n}$).
Since we have E$_{n}$ = nE the denominator is n!.

This was all for the tridiagonal case. Looking at Table 2 for the
pentadiagonal case we see that the coefficient of the v$^{n}$is 1/n!.
This is to be expected since, if we set w to be zero we get back to
the tridiagonal case {[}2{]}. There it was shown, and repeated in
Sec 3, that by considering the energy denominators, one gets this factor.

We can similarly get the other coefficients by considering all possible
paths to get to the nth state.. For example take the second term for
n=6, namely -v$^{4}$w/48. To evaluate the various energy denominators
we have to consider all possible arrangements and the corresponding
energy denominators.

.

vvvvw: 1{*}1/2{*}1/3{*}1/4{*}1/6 =1/144

vvvwv: 1{*}1/2{*}1/3{*}1/5{*}1/6 =1/180

vvwvv: 1{*}1/2{*}1/4{*}1/5{*}1/6 = 1/240

vwvvv: 1{*}1/3{*}1/4{*}1/5{*}1/6 = 1/360

wvvvv: 1/2{*}1/3{*}1/4{*}1/5{*}1/6 = 1/720

sum=1/48.

The negative sign comes from the fact that for v$^{4}$ w we have
5 negative energy denominators.

As another example consider v$^{2}$w$^{3}$ term for n=8. There are
5!/(2!{*}3!) = 10 partitons.

vvwww (2{*}4{*}6{*}8)$^{-1}$=1/384

vwvww (3{*}4{*}6{*}8)$^{-1}$ =1/576

vwwvw (3{*}5{*}6{*}8)$^{-1}$ =1/720

vwwwv (3{*}5{*}7{*}8)$^{-1}$ =1/840

wvvww (2{*}3{*}4{*}6{*}8)$^{-1}$= 1/1152

wwvvw (2{*}4{*}5{*}6{*}8)$^{-1}$ =1/1920

wwwvv(2{*}4{*}6{*}7{*}8)$^{-1}$ =1/2688

wvwvw (2{*}3{*}5{*}6{*}8)$^{-1}$=1/1440

wvwwv (2{*}3{*}5{*}7{*}8)$^{-1}$ =1/1680

wwvwv (2{*}4{*}5{*}7{*}8)$^{-1}$ =1/2240

sum= 1/96

.

\section{A comparison of the weak coupling and exact results}

.%
\noindent\begin{minipage}[t]{1\columnwidth}%
Table 3: Components a$_{n}$ forthe lowest eigenvector (ground state
wave function) for the pentaigonal matrix: Weak Coupling vs. Exact
. %
\end{minipage}

.%
\begin{tabular}{|c|c|c|}
\hline 
v=0.1  & w=0 & \tabularnewline
\hline 
\hline 
n & Weak & Exact\tabularnewline
\hline 
0 & 1 & 0.951\tabularnewline
\hline 
1 & -0.1 & -0.9902E-1\tabularnewline
\hline 
2 & 0.5000E-2 & 0.4934E-2\tabularnewline
\hline 
3 & -0.1667E-3 & -0.1641E-3\tabularnewline
\hline 
4 & 0.4167E-5 & 0.4094E-5\tabularnewline
\hline 
5 & -0.8333E-7 & -0.8174E-7\tabularnewline
\hline 
6 & 0.1389E-8 & 0.1360E-8\tabularnewline
\hline 
7 & -0.1984E-10 & -0.1941E-10\tabularnewline
\hline 
8 & 0.2480E-12 & 0.2435E-12\tabularnewline
\hline 
9 & -0.2756E-14 & -0.2690E-14\tabularnewline
\hline 
10 & 0.2756E-16 & 0.2687E-16\tabularnewline
\hline 
\end{tabular}%
\begin{tabular}{|c|c|c|}
\hline 
v=0 &  w=0.1 & \tabularnewline
\hline 
\hline 
n & Weak & Exact\tabularnewline
\hline 
0 & 1.0 & 0.9975\tabularnewline
\hline 
1 & 0 & 0\tabularnewline
\hline 
2 & 0.5000E-1 & -0.4988E-1\tabularnewline
\hline 
3 & 0 & 0\tabularnewline
\hline 
4 & 0.1250E-2 & 0.1246E-2\tabularnewline
\hline 
5 & 0 & 0\tabularnewline
\hline 
6 & -0.2084E-4 & -0.2075E-4\tabularnewline
\hline 
7 & 0 & 0\tabularnewline
\hline 
8 & 0.2604E-6 & 0.2593E-6\tabularnewline
\hline 
9 & 0 & 0\tabularnewline
\hline 
10 & -0.2604E-8 & -0.2591E-8\tabularnewline
\hline 
\end{tabular}%
\begin{tabular}{|c|c|c|}
\hline 
v=0.1  & w=0.1 & \tabularnewline
\hline 
\hline 
n & Weak & Exact\tabularnewline
\hline 
0 & 1.0 & 0.9941\tabularnewline
\hline 
1 & -0.1 & -0.9410E-1\tabularnewline
\hline 
2 & -0.4500E-1 & -0.4499E-1\tabularnewline
\hline 
3 & 0.4833E-2 & 0.4585E-2\tabularnewline
\hline 
4 & 0.1004E-2 & 0.1010E-2\tabularnewline
\hline 
5 & -0.1168E-3 & -0.1132E-2\tabularnewline
\hline 
6 & -0.1479E-4 & -0.1497E-4\tabularnewline
\hline 
7 & 0.1879E-5 & 0.1798E-5\tabularnewline
\hline 
8 & 0.1666E-6 & 0.1647E-6\tabularnewline
\hline 
9 & -0.2267E-7 & -0.2176E-7\tabularnewline
\hline 
10 & -0.1387E-8 & -0.1427E-8\tabularnewline
\hline 
\end{tabular}%
\begin{tabular}{|c|c|c|}
\hline 
v=0.1  & w =- 0.1 & \tabularnewline
\hline 
\hline 
n & Weak & Exact\tabularnewline
\hline 
0 & 1 & 0.9930\tabularnewline
\hline 
1 & -0.1 & -0.1036\tabularnewline
\hline 
2 & 0.5500E-1 & 0.5474E-1\tabularnewline
\hline 
3 & -0.5167E-1 & -0.5306E-1\tabularnewline
\hline 
4 & 0.1504E-2 & 0.1499E-2\tabularnewline
\hline 
5 & -0.1334E-3 & -0.1363E-3\tabularnewline
\hline 
6 & 0.2729E-4 & 0.2723E-4\tabularnewline
\hline 
7 & -0.2296E-5 & -0.2336E-5\tabularnewline
\hline 
8 & 0.3751E-6 & 0.3693E-6\tabularnewline
\hline 
9 & -0.2962E-7 & -0.3005E-7\tabularnewline
\hline 
10 & 0.3395E-8 & 0.3987E-8\tabularnewline
\hline 
\end{tabular}

.

. We see that for all the cases considered there the weak coupling
(or asymptotic) results are very close the exact results.Of course we
chose v/E and w/E to be 0.1 ,0 or -0.1.We can ask , where will the
asymptotic results no longer be valid? Obviously not at v=1. In the
weak coupling limit if v=1 then in all cases of table 3 we would have
that a$_{0}$ =1 and a$_{1}$=-1. If nothing else it would mean that
the eigenvector is not normalized to one.

We can also use table 3 see the trends of varying v and w. For v=0
w=1 the odd a$_{n}$vanish. This is because, starting from the ground
state one has to jump by 2 units to get a non-zero contribution. Relative
to the case v=0.1, w=0 we see that the non-zero w's in case 3 and
4 lead to an enhancement of the lower components a$_{n}$ . For example

with v=0.1 w=0 a$_{10}$= 0. 2686 E-15 but the values change to -0.1427E-8
for v=0.1 w=0.1 , and to 0.3987 E-8 for v=0.1 w=-0.1. 

.

\section{Added remarks}

Another point , there are several papers {[}6-10{]} where instead
of magnetic dipole excitations the authors deal with magnetic(and
electric) dipole decay. The latter is not the inverse of what we are
doing because in our case we always start from a certain state e.g.
the ground state and go directly to an excited state, whereas in {[}6-10{]}
there is, following an (n,$\gamma$) capture, a cascade of $\gamma$'s
to several intermediate states before reaching the ground state. With
out simplified interaction and simplifies model we found that the
cascade distribution of the average B(M1) strength was the similar in shape as the
excitation distribution for the total B(M1) strength. However the magnitudes are different as seen in the tables. Admittedly the association of our simplified models with the realistic models needs
a lot more work but we feel that these efforts are worth pursuing.In
this work we are trying to ascertain the simple properties of a more
complex matrix then  we previously studied in depth( pentadiagonal vs.
tridiagonal) and we hope to continued our 2 pronged attack ,by considering
realisitic calculations as in ref {[}5{]} and schematic ones as in
refs{[}1-4{]}.

\end{document}